\documentclass[prl,twocolumn, aps,superscriptaddress,floatfix,nofootinbib]{revtex4}

\usepackage{graphicx}
\usepackage{verbatim}
\usepackage{mathrsfs}
\usepackage{amsmath,amsfonts,amssymb}
\usepackage{graphicx}

\def\3{2.8in}    
\def\2{2.5in}
\def\4{3.0in}

\def \beq {\begin{equation}}
\def \eeq {\end{equation}}
\begin{document}

\title{Lifshitz transition and van Hove singularity in a topological Dirac semimetal}

\author{Su-Yang~Xu*}
\affiliation {Laboratory for Topological Quantum Matter and Spectroscopy (B7), Department of Physics, Princeton University, Princeton, New Jersey 08544, USA}

\author{Chang~Liu*}
\affiliation {Laboratory for Topological Quantum Matter and Spectroscopy (B7), Department of Physics, Princeton University, Princeton, New Jersey 08544, USA}

\author{I.~Belopolski}
\affiliation {Laboratory for Topological Quantum Matter and Spectroscopy (B7), Department of Physics, Princeton University, Princeton, New Jersey 08544, USA}

\author{S.~K.~Kushwaha}
\affiliation {Department of Chemistry, Princeton University, Princeton, New Jersey 08544, USA}
\author{R.~Sankar} \affiliation{Center for Condensed Matter Sciences, National Taiwan University, Taipei 10617, Taiwan}

\author{J.~W.~Krizan}
\affiliation {Department of Chemistry, Princeton University, Princeton, New Jersey 08544, USA}

\author{T.-R.~Chang}
\affiliation{Department of Physics, National Tsing Hua University, Hsinchu 30013, Taiwan}

\author{C.~M.~Polley}
\affiliation {MAX-lab, Lund University, S-22100 Lund, Sweden}

\author{J.~Adell}
\affiliation {MAX-lab, Lund University, S-22100 Lund, Sweden}

\author{T.~Balasubramanian}
\affiliation {MAX-lab, Lund University, S-22100 Lund, Sweden}

\author{K.~Miyamoto}
\affiliation {Hiroshima Synchrotron Radiation Center, Hiroshima University, 2-313 Kagamiyama, Higashi-Hiroshima 739-0046, Japan}

\author{N.~Alidoust}
\affiliation {Laboratory for Topological Quantum Matter and Spectroscopy (B7), Department of Physics, Princeton University, Princeton, New Jersey 08544, USA}

\author{Guang~Bian}
\affiliation {Laboratory for Topological Quantum Matter and Spectroscopy (B7), Department of Physics, Princeton University, Princeton, New Jersey 08544, USA}

\author{M.~Neupane}
\affiliation {Laboratory for Topological Quantum Matter and Spectroscopy (B7), Department of Physics, Princeton University, Princeton, New Jersey 08544, USA}

\author{H.-T.~Jeng}
\affiliation{Department of Physics, National Tsing Hua University, Hsinchu 30013, Taiwan}
\affiliation{Institute of Physics, Academia Sinica, Taipei 11529, Taiwan}

\author{C.-Y. Huang}
\affiliation{Department of Physics, National Sun Yat-Sen University, Kaohsiung 804, Taiwan}

\author{W.-F. Tsai}
\affiliation{Department of Physics, National Sun Yat-Sen University, Kaohsiung 804, Taiwan}

\author{T.~Okuda}
\affiliation {Hiroshima Synchrotron Radiation Center, Hiroshima University, 2-313 Kagamiyama, Higashi-Hiroshima 739-0046, Japan}

\author{F.~C.~Chou} \affiliation{Center for Condensed Matter Sciences, National Taiwan University, Taipei 10617, Taiwan}
\author{R.~J.~Cava}
\affiliation {Department of Chemistry, Princeton University, Princeton, New Jersey 08544, USA}
\author{A.~Bansil}
\affiliation {Department of Physics, Northeastern University, Boston, Massachusetts 02115, USA}
\author{H.~Lin}
\affiliation {Graphene Research Centre and Department of Physics, National University of Singapore 11754, Singapore}

\author{M.~Z.~Hasan}
\affiliation {Laboratory for Topological Quantum Matter and Spectroscopy (B7), Department of Physics, Princeton University, Princeton, New Jersey 08544, USA}


\date{\today}

\begin{abstract}
A topological Dirac semimetal is a novel state of quantum matter which has recently attracted much attention as an apparent 3D version of graphene. In this paper, we report results on the electronic structure of the 3D Dirac semimetal Na$_3$Bi at a surface that reveals its nontrivial groundstate. Our studies, for the first time, reveal that the two 3D Dirac cones go through a topological change in the constant energy contour as a function of the binding energy, featuring a Lifshitz point, which is missing in a strict 3D analog of graphene (in other words, Na3Bi is not a true analog of 3D graphene). Our results identify the first example of a band saddle point singularity in 3D Dirac materials. This is in contrast to its 2D analogs such as graphene and the helical Dirac surface states of a topological insulator. The observation of multiple Dirac nodes in Na$_3$Bi connecting via a Lifshitz point along its crystalline rotational axis away from the Kramers point serves as a decisive signature for the symmetry-protected nature of the Dirac semimetal's topological \textit{bulk} groundstate.
\end{abstract}

\maketitle
Three-dimensional (3D) Dirac semimetals are materials whose bulk band structure features 3D Dirac fermion quasi-particles \cite{3D_Dirac, Weyl, Volovik2003, Dirac_3D, Dirac_semi, Dai, Nagaosa, Sanvito, RMP, Zhang_RMP, Suyang, David Nature BiSb, Matthew Nature physics BiSe, NbSe2}. It was theoretically predicted that a 3D Dirac semimetal state can be realized at the critical point of the topological phase transition between a band insulator and a 3D topological insulator in a space inversion symmetric bulk crystal \cite{3D_Dirac}. The Dirac semimetal state realized in this way features a single 3D Dirac cone at the Kramers point where the bulk band inversion occurs \cite{3D_Dirac}. This type of 3D Dirac semimetal has been realized and observed in multiple material systems such as BiTl(S$_{0.5}$Se$_{0.5}$)$_2$ and (Bi$_{0.94}$In$_{0.06}$)$_2$Se$_3$ \cite{Suyang, Oh}. However, this approach requires fine-tuning the chemical doping/alloying composition, which is difficult to control and also introduces chemical disorders limiting the mobility of the system. In 2012 and 2013, numerical band calculations predicted 3D Dirac semimetal states in two stoichiometric materials, Na$_3$Bi and Cd$_3$As$_2$ \cite{Dirac_semi, Dai}. Unlike the fine-tuned case, the Dirac semimetal states in Na$_3$Bi and Cd$_3$As$_2$ show a pair of bulk Dirac nodes that are located along the rotation axis away from the Kramers point in calculation \cite{Dirac_semi, Dai}. This new type of Dirac semimetal with multiple bulk Dirac nodes arises from the protection by additional crystalline rotational symmetries \cite{Dirac_semi, Dai, Nagaosa}. They are believed to be stable within a range of material parameter space and robust to finite disorder or chemical doping. As a result, symmetry-protected Dirac semimetal states have soon attracted much interest \cite{Dirac_semi, Dai, Nagaosa}. Following these two band structure predictions, several photoemission experiments studied the (001) surface electronic structure of Na$_3$Bi and Cd$_3$As$_2$ \cite{Chen_Na3Bi, CdAs_Hasan, CdAs_Cava}. Although 3D Dirac cones have been observed in both materials \cite{Chen_Na3Bi, CdAs_Hasan, CdAs_Cava}, it was found difficult to resolve the existence of two bulk Dirac cones and their overlap/interplay in momentum space when studying the (001) surface. Therefore, the critically important unique features of the electronic structure of symmetry-protected Dirac semimetals have not been unambiguously probed in experiments. Revealing the true electronic groundstate is fundamental for future works on any new quantum materials, either theoretically or experimentally. This is even more important for a symmetry protected Dirac semimetal such as Na$_3$Bi because it is the phenomenon of multiple bulk Dirac nodes away from the Kramers point along the rotation axis that distinguishes this new type of symmetry protected Dirac semimetal state from the former (fine-tuned) type that is achieved by fine-tuning \cite{Nagaosa}.


In this Letter, we solve this problem by studying the electronic structure of Na$_3$Bi at a different surface - the (100) surface by angle-resolved photoemission spectroscopy (ARPES). The (001) and the (100) surfaces correspond to the ($0001$) and the ($10\bar{1}0$) surfaces in the hexagonal four-index scheme.  Our studies, for the first time, reveal unambiguously that the electronic groundstate Dirac semimetal material Na$_3$Bi consist of two 3D Dirac cones and that these two Dirac cones go through a Lifshitz transition, a topological change in the constant energy contour \cite{Lifshitz}, which is missing in a strict 3D analog of graphene. Our results identify the first example of van Hove singularity in a 3D Dirac system, and therefore pave the way for studying unusual phenomena related to van Hove singularities in these novel materials. These observations are made possible because we were able to access the (100) surface. At the (001) surface (as the case in previous studies \cite{Chen_Na3Bi, CdAs_Hasan, CdAs_Cava}), the experiments were limited by the out-of-plane momentum resolution of ARPES probes.

\begin{figure}
\centering
\includegraphics[width=8.5cm]{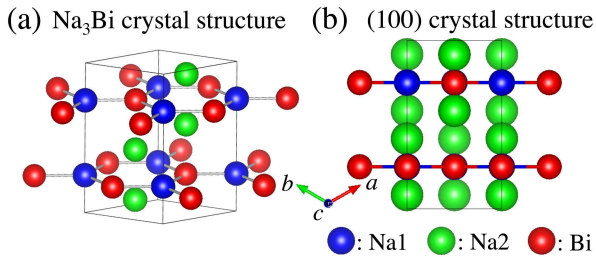}
\caption{\textbf{Crystal structure and multiple Dirac nodes in Na$_3$Bi.} (\textbf{Characterization of Na$_3$Bi system.} (a) Crystal structure of Na$_3$Bi. The two Na sites and Bi atoms are marked with different colors. (b) Projected crystal structure at the (100) surface. (d) First-principles bulk band structure calculation. (d) Structure of bulk and surface Brillouin zone at (001) and (100) surfaces. Bulk Dirac nodes are marked by green crosses. Panels (c,d) are adapted from Ref. \cite{Dirac_semi}.}
\end{figure}

\begin{figure}
\centering
\includegraphics[width=8.5cm]{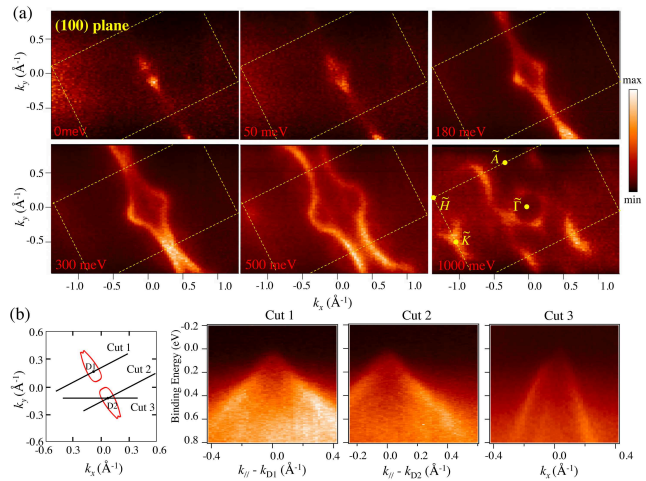}
\caption{\textbf{ARPES observation of multiple bulk Dirac cones.} (a) ARPES constant energy maps obtained at the (100) surface using incident photon energy of 58 eV. Yellow dashed boxes indicate the surface Brillouin zone; notations of high symmetry points are shown at the $E_\textrm{B} = 1000$ meV panel; binding energies are marked in red on the lower left corner of each panel. (b) Selected ARPES \textit{k}-\textit{E} cuts, directions and rough $k$-ranges of the cuts are indicated by black solid lines in the left panel.}
\end{figure}

\begin{figure*}
\centering
\includegraphics[width=15cm]{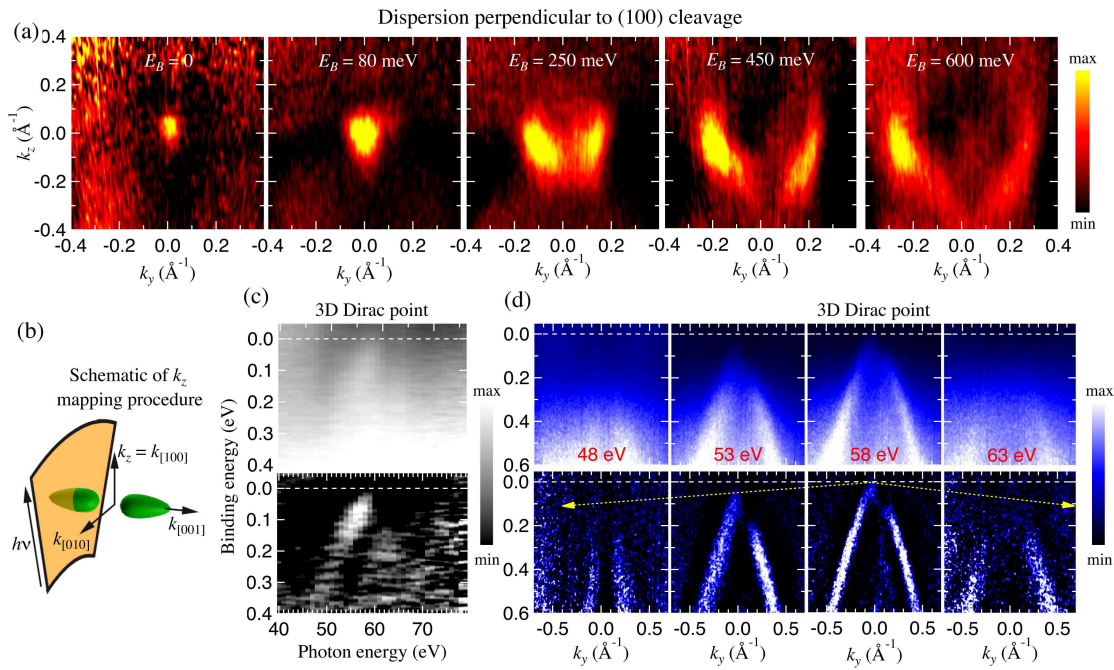}
\caption{\textbf{Out-of-plane dispersion of the bulk Dirac fermions.} \textbf{a}, ARPES constant energy maps along the $k_y$-$k_z$ plane. Binding energies are marked on top of each panel. The $k_z$ [100] axis is built by measuring ARPES $k$-$E$ maps with photon energies ranging from 40 to 76 eV. \textbf{b}, Schematic diagram of the $k_z$ mapping procedure. Since the cleaving surface is (100), $k_z$ corresponds to the [100] direction parallel to the crystallographic $a$ axis. \textbf{c}, Linear dispersion along the third dimension ($k_z$). Probing with different photon energies, the top of one of the bulk Dirac cones locate at different binding energies, constructing the linear $k_z$ dispersion. \textbf{d}, Linear $k_z$ dispersion seen from in-plane $k$-$E$ cuts. It is clear from the data that the bulk Dirac cone does not touch the Fermi level except for $h\nu=58$ eV, consistent with the three dimensional Dirac dispersion. For Panels \textbf{c} and \textbf{d}, top row presents raw data, Bottom row shows second derivative results along the momentum distribution curves.}\label{kz}
\end{figure*}

\begin{figure}
\centering
\includegraphics[width=8.5cm]{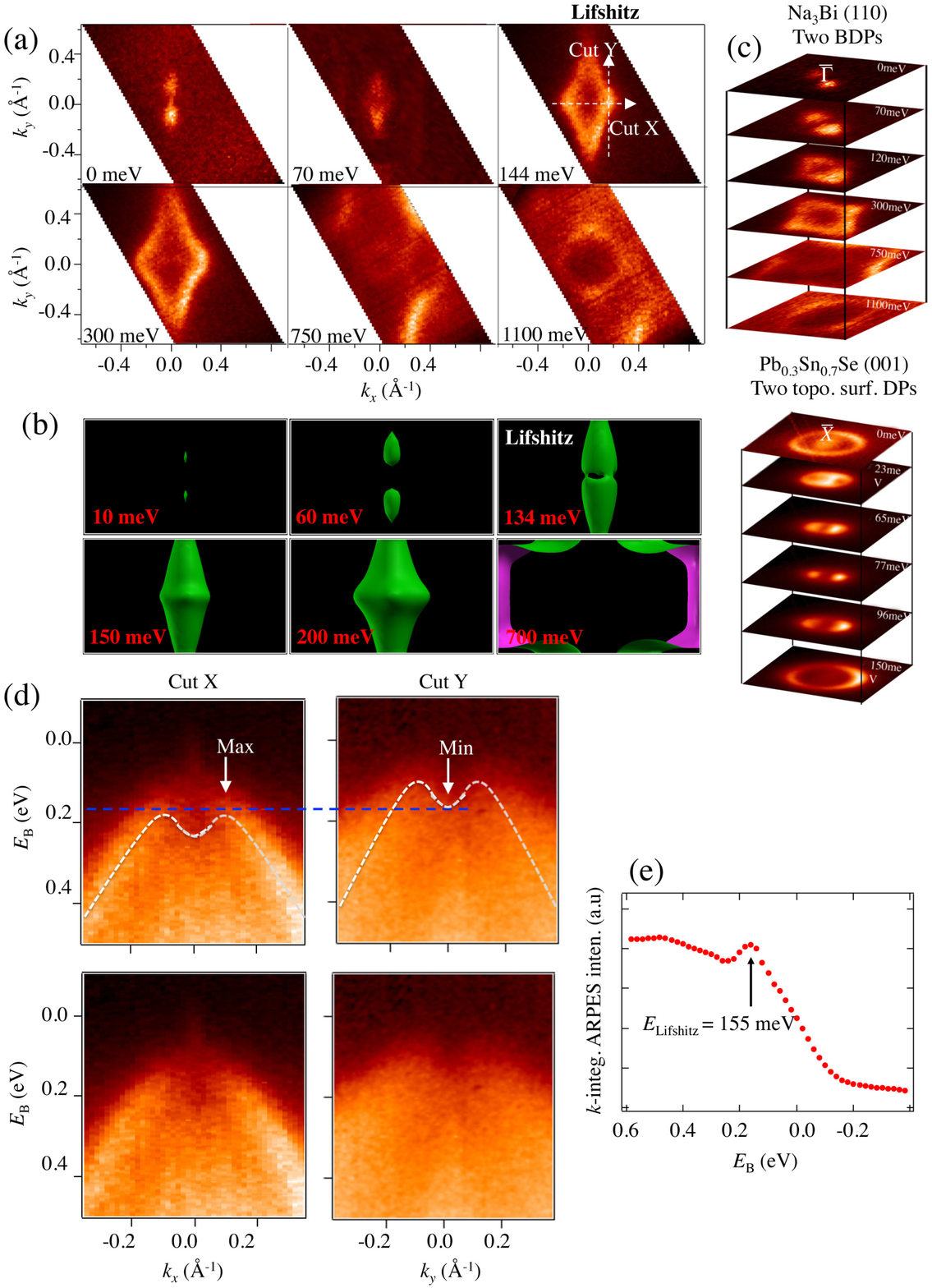}
\caption{\textbf{Lifshitz transition and saddle point singularity in the bulk Dirac cones.} (a) Lifshitz transition observed in ARPES constant energy maps. At $E_\textrm{B} \sim 144$ meV the two cones touch and hybridize to form a diamond shaped single band contour, marking the Lifshitz transition. (b) Theoretical 3D constant energy maps at various binding energies, for comparison with Panel (a). Inset in the 134 meV panel shows band projections onto the (100) surface at the Lifshitz energy. CEM: constant energy map. (c) Lifshitz transition of \textit{bulk} Dirac fermions in Na$_3$Bi (top panel), compared with the Lifshitz transition of \textit{surface} Dirac gas in a topological crystalline insulator Pb$_{0.7}$Sn$_{0.3}$Se (bottom panel). (d) ARPES $k$-$E$ maps along Cut X and Cut Y defined in Panel (a) with and without the dashed lines. The dashed lines are guides to the eyes, which are drawn to only show the trend of the band dispersion at a qualitative level. (e) Saddle point singularity observed from an anomalous increase of local density of states. The peak of integrated ARPES intensity marks the Lifshitz transition at $E_\textrm{B} \sim 155$ meV. The EDC is obtained by integrating the ARPES intensity at a $(\pm0.1$ $\textrm{\AA}^{-1}, \pm0.1$ $\textrm{\AA}^{-1})$ window in the vicinity of the saddle point, which is the point where Cut X and Cut Y cross each other in panel (a).}
\end{figure}

High quality single crystals of Na$_3$Bi were grown by a method reported in Ref. \cite{Brauer}. Ultraviolet spin-integrated ARPES measurements were performed at the beamline I4 at the MAX-lab in Lund in Sweden and the ESPRESSO endstation installed at the beamline-9B of the Hiroshima Synchrotron Radiation Center (HiSOR) in Hiroshima in Japan. The energy and momentum resolution was better than 30 meV and $1\%$ of the BZ for spin-integrated ARPES measurements at the beamline I4 at the MAX-lab and the ESPRESSO endstation at the beamline-9B of Hisor. Samples were cleaved \textit{in situ} under a vacuum condition better than $1 \times 10^{-10}$ Torr at all beamlines.  Since bismuth trisodium is air sensitive, argon-filled glove boxes with residual oxygen and water level less than 1 ppm were used in the entire preparation process. ARPES measurements were performed at liquid nitrogen temperatures in the I4 beamline at the MAX-lab, and at 10-20 K at the beamline 9B at HiSOR. Samples were found to be stable and without degradation for a typical measurement period of 24 hours. The first-principles calculations were based on the generalized gradient approximation (GGA) \cite{Perdew} using the full-potential projected augmented wave method as implemented in the VASP package. Experimental lattice parameters were adopted from Ref. \cite{Brauer}. The electronic structure of bulk bismuth trisodium was calculated using a $9\times5\times5$ Monkhorst-Pack $k$-mesh over the BZ with the inclusion of spin-orbit coupling.

Na$_3$Bi crystalizes in the hexagonal $P6_3/mmc$ crystal structure with $a=5.448$ $\textrm{\AA}$ and $c=9.655$ $\textrm{\AA}$ \cite{Na3Bi_crystal} ($a^{-1}/c^{-1}\simeq1.8$). It is a layered structure consisting of a honeycomb atomic plane where the Na and Bi atoms are situated on the A and B sites respectively, sandwiched by layers of Na atoms which form a triangular lattice [see Figs. 1(a,b)]. In the ionic limit, the bulk conduction and valence bands are expected to be the Na $3s$ and the Bi $6p_{x,y,z}$, respectively. However, first-principles calculation with spin-orbit interaction \cite{Dirac_semi} shows that the conduction and valence bands are inverted at the bulk BZ center $\Gamma$ point. Since the [001] axis ($\Gamma-A$ direction in momentum space) is a $C_3$ rotational axis, all electronic bands along the $\Gamma-A$ direction consist of states, which are eigenstates of the $C_3$ rotational operator. Interestingly, in Na$_3$Bi, the lowest conduction and valence bands that go through the band inversion at the $\Gamma$ point are found to have different $C_3$ eigenvalues \cite{Dirac_semi, Nagaosa}. This fact suggests that an energy gap cannot open between these two bands along the $\Gamma-A$ direction even after including spin-orbit interaction. Consequently, two bulk Dirac band crossings along the $A-\Gamma-A$ direction on the opposite sides of the $\Gamma$ point are shown in first-principles calculations [Fig. 1(c)]. Therefore, the two bulk Dirac points are expected to be located along the [001] axis \cite{Dirac_semi, Dai}, which is perpendicular to the (001) surface \cite{Chen_Na3Bi, CdAs_Hasan, CdAs_Cava}. Since the out-of-plane momentum resolution of ARPES is much limited, it is difficult to resolve the existence of two bulk Dirac cones and their overlap/interplay at the (001) surface. By contrast, they project onto different points in the (100) surface BZ  [Fig. 1(d)].

Fig. 2 shows our measured energy-momentum dispersion of the two bulk Dirac cones at the (100) surface. The ARPES constant energy maps are shown in Fig. 2(a). The surface BZ is found to be in a rectangular shape, which is consistent with the (100) surface termination. The length ratio between the long and short edge of the surface BZ (yellow boxes) is measured to be about $1.8 \simeq a^{-1}/c^{-1}$, also consistent with a (100) cleaving surface. Remarkably, at the Fermi level, our data clearly show that the Fermi surface consists of two unconnected band contours, which represent the two bulk Dirac cones. And the two bulk Dirac cones are observed to be equally spaced on the opposite sides of the $\Gamma$ point along the $A-\Gamma-A$ direction. A slight increase of binding energy up to 50 meV results in enlargement of the two Fermi pockets. At higher binding energies this two cones merge. The band contour further expands to form a neck region around the $\tilde{A}$ points (top of the 3D BZ). The chemical potential seems to be slightly below the energy of the two Dirac points, since the observed two separated pockets still has some finite area. Fig. 2(b) presents several typical ARPES $k$-$E$ maps along various cutting directions [black solid lines in the left panel of Fig. 2(b)]. A clear linear dispersion is seen for the bulk Dirac cones in all of these maps, establishing the existence of massless bulk quasiparticles. In Fig. 3 we study the band dispersion along the out-of-plane direction that is perpendicular to the (100) cleavage surface, by varying the incident photon energies of the synchrotron radiation. We expect the constant energy ($k_y-k_z$) maps to demonstrate an enlarging, circular-like band contour. From Fig. 3(a) one realizes that this is indeed the case. Moreover, the $E$-$k_z$ ($E$-$k_{[100]}$) cut shown in Fig. 3(c) is also consistent with the expected linear dispersion. In Fig. 3(d) the same fact is emphasized again by showing the $k_y$-$E$ maps for multiple photon energies. One realizes from Fig. 3(d) that the Dirac band touches the Fermi level only at $h\nu = 58$ eV.

Since the two bulk Dirac cones are located very close to one another in momentum space, they inevitably touch and hybridize with each other, giving rise to a topological change in the band contours, also known as a Lifshitz transition in the electronic structure. Fig. 4 demonstrates our systematic ARPES investigation on the Lifshitz transition as well as the associated saddle point band structure. Fig. 4(a) presents a set of ARPES constant energy maps with higher momentum resolution than the one shown in Fig. 2. Besides the two bulk Dirac cones presented already in Fig. 2, we emphasize here the change of band structure where the two unconnected Dirac band contours merge. Experimentally, this Lifshitz transition happens at a binding energy of $\sim144$ meV, at which the two hole-like bands transform to a diamond-shaped loop. At higher binding energies, this loop enlarges while another concentric contour starts to appear at the zone center. This trend of band evolution is nicely reproduced in first principles band calculations, as shown in Fig. 4(b) where a set of 3D ($k_{[100]}$, $k_{[010]}$, $k_{[001]}$) constant energy maps at selected binding energies are presented. The theoretical critical point of the Lifshitz transition is at$E_\textrm{B} = 134$ meV, only 10 meV away from the experimentally observed value. It can be seen in Fig. 4(a) that the two cones touch at two special momenta, ($k_x$, $k_y$) = ($\pm0.1$, 0) $\textrm{\AA}^{-1}$. To directly identify the saddle point band structure as a result of the Lifshitz transition in experiments, we center our ARPES detector at one of the special momenta at ($k_x$, $k_y$) = ($+0.1$, 0) $\textrm{\AA}^{-1}$ and study two $k$-$E$ maps, namely, Cuts X and Y as defined in the $3^{\mathrm{rd}}$ panel in Fig. 4(a). As shown in Fig. 4(d), the hybridized band reaches its local energy maximum at ($k_x$, $k_y$) = ($+0.1$, 0) $\textrm{\AA}^{-1}$ along $k_x$, while reaching a local energy minimum along $k_y$. This behavior defines a saddle point singularity of band structure. At Fig. 4(e), an increase of local density of state (DOS) is observed by integrating the ARPES intensity around the saddle point, consistent with a local DOS maximum in a saddle point type of van Hove singularity in three-dimensions. In Fig. 4(c) we compare and contrast the Lifshitz transition observed here with that observed in the surface states of a topological crystalline insulator Pb$_{0.7}$Sn$_{0.3}$Se. In the $n$-typed sample of Pb$_{0.7}$Sn$_{0.3}$Se, the two \textit{surface} Dirac cones hybridize at $E_\textrm{B} = 23$ and 96 meV. The difference between these two cases is that for Na$_3$Bi, bulk Dirac quasiparticles, instead of the surface Dirac electron gas in the case of TCI, merge and hybridize.

We note the following observations in our experiments: (i) we observe a pair of bulk Dirac nodes; (ii) they are located along the $C_3$ $k_{[001]}$ axis away from the $\Gamma$ point; (iii) the distance between the two in momentum space is only 0.2 $\textrm{\AA}^{-1}$; (iv) the Lifshitz transition and saddle point singularity between the two are resolved. These observations are made possible because we studied (100) surface. At the (001) surface \cite{Chen_Na3Bi, CdAs_Hasan, CdAs_Cava}, it is difficult due to the fact that the out-of-plane momentum resolution of ARPES is quite limited. We discuss the connection and distinction between the saddle point type of van Hove singularities in 2D and 3D band structures. Saddle point singularities in 2D Dirac electron gas have been extensively studied in twisted bi-layer graphene \cite{Kim, Geim} and the surfaces of topological crystalline insulators \cite{Saddle}. In 2D, the DOS at the energy of the singularity diverges logarithmically fashion leading to electronic instability and correlated physics. In contrast, in 3D, the DOS itself reaches a finite maximum, but the derivative of the DOS does diverge. However, unconventional correlation physics can still arise if the bulk band structure is in a good quasi-2D approximation, such as the case in high-$T_c$ copper superconductors \cite{Quasi-2D}. Despite this difference, our results have, for the first time, identify an example of band structure singularity in 3D Dirac materials. Moreover, we note that strictly speaking, Lifshitz transitions are topological changes of the Fermi surface upon changing some external parameter (doping, pressure, temperature, etc). Thus to achieve such a Lifshitz transition and to access the singularity at the Fermi level, one needs to dope Na$_3$Bi to tune the chemical potential. A recent paper showed that controlling the chemical potential is indeed possible by finely adjusting the growth conditions of Na$_3$Bi \cite{Satya}. Furthermore, we note that a very recent theory work has classified all possible 3D Dirac semimetal states \cite{Nagaosa}: theoretically, there are three classes of Dirac semimetals \cite{Nagaosa}, namely the (trivial) Dirac semimetal realized at the critical point of a topological phase transition, the (topological) Dirac semimetal state with a pair of Dirac nodes protected by the rotational symmetry, and the single-Dirac-cone Dirac semimetal state at the Kramers point protected by rotational symmetry. Our unambiguous observation of a pair of bulk Dirac nodes along the crystalline rotational axis away from the Kramers $\Gamma$ point serve as a signature for the symmetry-protected nature of the Dirac semimetal state, and, for the first time, revealed that the groundstate of Na$_3$Bi is a \textit{nontrivial} 3D generalization of graphene.

\end{document}